%% file: main.tex
\documentclass{fun_article_template}
\title{Cascades towards noise-induced transitions on networks revealed using information flows}

\usepackage{authblk}    %
\usepackage{abstract}   %
\usepackage{lettrine}   %
\usepackage{lipsum}     %

\author{
	Casper van Elteren\textsuperscript{1,2}, Rick Quax \textsuperscript{1,2}, and Peter M.A. Sloot\textsuperscript{1,2,3} %
	\\% Space before institutions
	\textsuperscript{1}\institution{Computational Science Lab Amsterdam}\\ %
	\textsuperscript{2}\institution{Institute for Advanced Study Amsterdam}\\ %
	\textsuperscript{3}\institution{Complexity Science Hub Viennna} %
}

\renewenvironment{onecolabstract}
  {\begin{center}
   \bfseries\vspace{-8em}\vspace{0pt} %
   \end{center}\bfseries\quotation}
  {\endquotation}

\newcommand*{\addFileDependency}[1]{%
\typeout{(#1)}%
\@addtofilelist{#1}
\IfFileExists{#1}{}{\typeout{No file #1.}}
}

\makeatother

\date{}

\begin{document}
\twocolumn[
  \begin{@twocolumnfalse}
    \maketitle
    \begin{onecolabstract}
\lettrineabstract{Complex networks, from neuronal assemblies to social systems, can exhibit abrupt, system-wide transitions without external forcing. These endogenously generated ``noise-induced transitions'' emerge from the intricate interplay between network structure and local dynamics, yet their underlying mechanisms remain elusive. Our study unveils two critical roles that nodes play in catalyzing these transitions within dynamical networks governed by the Boltzmann-Gibbs distribution. We introduce the concept of ``initiator nodes'', which absorb and propagate short-lived fluctuations, temporarily destabilizing their neighbors. This process initiates a domino effect, where the stability of a node inversely correlates with the number of destabilized neighbors required to tip it. As the system approaches a tipping point, we identify ``stabilizer nodes'' that encode the system's long-term memory, ultimately reversing the domino effect and settling the network into a new stable attractor. Through targeted interventions, we demonstrate how these roles can be manipulated to either promote or inhibit systemic transitions. Our findings provide a novel framework for understanding and potentially controlling endogenously generated metastable behavior in complex networks. This approach opens new avenues for predicting and managing critical transitions in diverse fields, from neuroscience to social dynamics and beyond.}
    \end{onecolabstract}
      \end{@twocolumnfalse}
]

\section{Introduction}
\label{sec:orgd6a1d62}
Multistability, a fundamental characteristic of complex systems \cite{Ladyman2013, vanNes2016}, describes the capacity of a system to occupy multiple stable states and transition between them. This phenomenon is ubiquitous, manifesting in diverse domains from neural networks \cite{Kandel2000, Fries2015} to opinion dynamics \cite{Galam2020} and ecosystems \cite{Wunderling2021}. While state transitions are often attributed to external perturbations, we propose a novel perspective: in networked systems, noise-induced transitions can occur endogenously. These transitions emerge from local interactions that cascade through the network, triggering large-scale regime shifts in a process we term the ``domino effect''. This mechanism offers a new understanding of how complex systems can dramatically recon figure without external forcing, challenging traditional views on system stability and change.

In nonlinear systems, such as interconnected neurons, noise plays a fundamental role in facilitating transitions between attractor states \cite{Beggs2012, Mitchell1993, Forgoston2018a}. It enables the exploration of larger state spaces, allowing systems to escape local minima \cite{Czaplicka2013, Nicolis2016}. While multistability has historically been studied from an equilibrium perspective \cite{McNamara1989, Kramers1940, Czaplicka2013a}, recent research has revealed how network structure fundamentally affects the stability and transitions of complex systems \cite{Harush2017a, Gao2016, Dong2021, Liu2021}.

Our study addresses a critical gap in understanding noise-induced transitions in networked dynamical systems out of equilibrium. We focus on systems where each node's state evolves according to the Boltzmann-Gibbs distribution, a framework applicable to various phenomena including neural dynamics \cite{Hopfield1982b}, opinion formation, and ferromagnetic spins \cite{Glauber1963}.

We introduce two novel concepts: \emph{initiator} nodes that propagate noise and destabilize the system, and \emph{stabilizing} nodes that maintain metastable states. To quantify the impact of short-term and long-term correlations in these transitions, we propose two information-theoretic measures: integrated mutual information and asymptotic information. These metrics, computable from observational data, provide powerful tools for analyzing metastable dynamics across different time scales.

\begin{figure*}
\centering
\includegraphics[width=.9\linewidth]{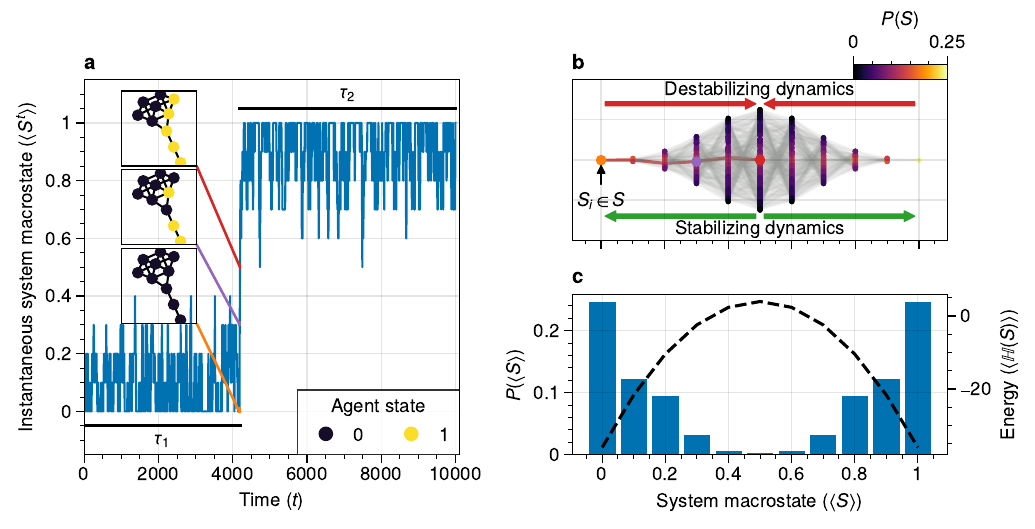}
\caption{\label{fig:introduction}A dynamical network governed by kinetic Ising dynamics produces multistable behavior. (a) A typical trajectory is shown for a kite network for which each node is governed by the Ising dynamics with \(\beta \approx 0.534\). The panels show system configurations \(S_i \in S\) as the system approaches the tipping point (orange to purple to red). For the system to transition between attractor states, it has to cross an energy barrier (c). (b) The dynamics of the system can be represented as a graph. Each node represents a system configuration \(S_i \in S\) such as depicted in (a). The probability for a particular system configuration \(p(S)\) is indicated with a color; some states are more likely than others. The trajectory from (a) is visualized. Dynamics that move towards the tipping point (midline) destabilize the system, whereas moving away from the tipping point are stabilizing dynamics. (c) The stationary distribution of the system is bistable. Crossing the tipping point requires crossing a high energy states (dashed line). Transitions between the attractor states are infrequent and rare. For more information on the numerical simulations see \cref{method:appendix}.}
\end{figure*}

Integrated mutual information captures the transient destabilization of the system, revealing the role of initiator nodes in triggering systemic transitions. Asymptotic information, on the other hand, quantifies the long-term memory encoded by stabilizer nodes, which ultimately reverse the domino effect and settle the network into a new stable attractor. By manipulating these roles, we demonstrate how targeted interventions can either promote or inhibit systemic transitions, offering a new approach to controlling critical transitions in complex networks.

Our computational method uncovers a network percolation process that facilitates noise-induced transitions without external parameter changes, offering a fresh perspective on tipping points in complex networks \cite{Lenton2024, Peng2019, Bury2021, DOrsogna2015a}. This approach bridges the gap between local equilibrium dynamics and global system behavior, providing insights into how network structure influences systemic transitions \cite{Harush2017a, Gao2016, Wunderling2020, Yang2017, Yang2017a}.

By revealing the the domino-like mechanisms of endogenous state transitions, our work has broad implications for predicting and potentially controlling critical transitions in diverse complex systems. From enhancing brain plasticity to anticipating ecosystem shifts, this framework provides a foundation for understanding and managing multistability in an interconnected world.

\section{Methods}
\label{sec:org6f0b033}
Our study focuses on dynamical systems where the state transitions of individual nodes are governed by the Boltzmann-Gibbs distribution. This distribution, fundamental in statistical mechanics, provides a probabilistic framework for describing the behavior of systems in thermal equilibrium. In our context, it determines the likelihood of a node transitioning from one state to another based on the energy difference between states and a global noise parameter. Specifically, the probability of a node transitioning from state $s_i$ to state $s_i'$ is given by:
\begin{equation}
P(s_i \to s_i') = \frac{1}{1 + \exp(-\beta \Delta E(s_i, s_i'))},
\end{equation}
\noindent where $\Delta E(s_i, s_i')$ represents the energy difference for the state transition, and $\beta$ is the inverse temperature or noise parameter. This formulation captures the essence of how local interactions and global noise influence state changes in our networked system. Higher values of $\beta$ correspond to lower noise levels, leading to more deterministic behavior, while lower $\beta$ values introduce more randomness into the system's dynamics. This framework allows us to model a wide range of phenomena, from neural activity to opinion dynamics, within a consistent mathematical structure.

Fluctuations and their correlations at time \(\tau\) are captured using  Shannon's  mutual information  \cite{Cover2005}  shared between  a   node's state ($s_i^t$) at time $t$  and   the  entire  future   system  state ($S^{t+\tau}$), \(I(s_i^{\tau}  : S^{\tau  +  t})\). The  time lag  \(t\)  is used  to analyze two key  features of information flows  of a system: the area  under the  curve (AUC) of  short-term information, and sustained level of long term information.

The contribution  of a  node to the  dynamics of  the system
will differ depending on the  network connectivity of a node
(\cref{fig:maj_flip})    \cite{vanElteren2022,Quax2013}.   The
total  amount  of  fluctuations shared  between  the  node's
current state and the  system's short-term future trajectory
is computed as the integrated mutual information

\begin{equation}
\label{eq:adj_imi}
\begin{split}
\mu(s_i) = \sum_{t = 0}^\infty (I(s_i^{\tau} : S^{\tau + t}) - \omega_{s_i}) \Delta t.
\end{split}
\end{equation}

Intuitively,  \(\mu(s_i)\)  represents   a  combination  of  the intensity and duration of the short-term fluctuations on the (transient) system dynamics \cite{vanElteren2022}. It reflects how much of the node state is in the ``working memory'' of the system.

The term $\omega(s_i) \in \mathbb{R}_{\ge 0}$ represents the system's long-term memory. As the system transitions between stable points, short-lived correlations evolve into longer-lasting ones, particularly among less dynamic nodes. When $\omega(s_i)$ is positive, it indicates a separation of time scales: ephemeral correlations dissipate, giving way to slower, more persistent fluctuations. These enduring fluctuations reflect the multiple attractor states accessible to the system, with less dynamic nodes becoming more aligned with future system states.

Near a stable attractor, the system primarily generates short-lived fluctuations. However, as it approaches a tipping point, longer-lasting correlations emerge. These persistent correlations facilitate the system's transition from one stable attractor to another, much like repeated nudges eventually push a ball over a hill. The asymptotic information, $\omega(s_i)$, quantifies this transition potential. Higher values of $\omega(s_i)$ indicate a greater likelihood of state transition, with the exact value reflecting each node's contribution to the tipping behavior.

Asymptotic information distinguishes itself from other early warning signals—such as increased autocorrelation, critical slowing down captured by Fisher information, changes in skewness or kurtosis, and increased variance—by specifically measuring the system's long-term memory and temporal correlation structure. While entropy captures the overall uncertainty or disorder in a system at a given moment, and mutual information quantifies the shared information between components at a particular time, asymptotic information focuses on the persistence of correlations over extended time periods. It reveals how past states influence future configurations, capturing aspects of the system's dynamics that are not explained by instantaneous or short-term pairwise measures.

Using these information features,  each node can be assigned to a different \emph{role} based on their contribution to the metastable transition. We denote nodes with short-lived  correlations as \emph{initiators} pushing nodes towards a tipping point. In contrast, nodes with longer-lived correlations are referred to as \emph{stabilizers}. For these nodes, their dynamics are less affected by short-lived correlations, and they require a higher mixing state for to transition from one state to another. The role assignment will be further discussed in  \ref{sec:roles}.

We compute information flows using exact calculations on a randomly generated connected graph of $n=10$ nodes. The states are grouped based on their distance to the tipping point, defined as the energy barrier between two locally stable states. For the Ising model, this corresponds to the collection of states where $\langle S \rangle = 0.5$. We evaluate the conditional distribution up to $\tau = 300$ time steps.

This computational process scales exponentially with the number of nodes, $O(n) = 2^n$, which limits its applicability to large-scale systems without employing variable reduction techniques such as coarse-graining. Extending this analysis to larger systems will be the focus of future research.

For detailed replication instructions, please refer to \cref{sec:org854db8e}.

\section{Results}
\begin{figure*}[bt]
\centering
\includegraphics[width=.9\linewidth]{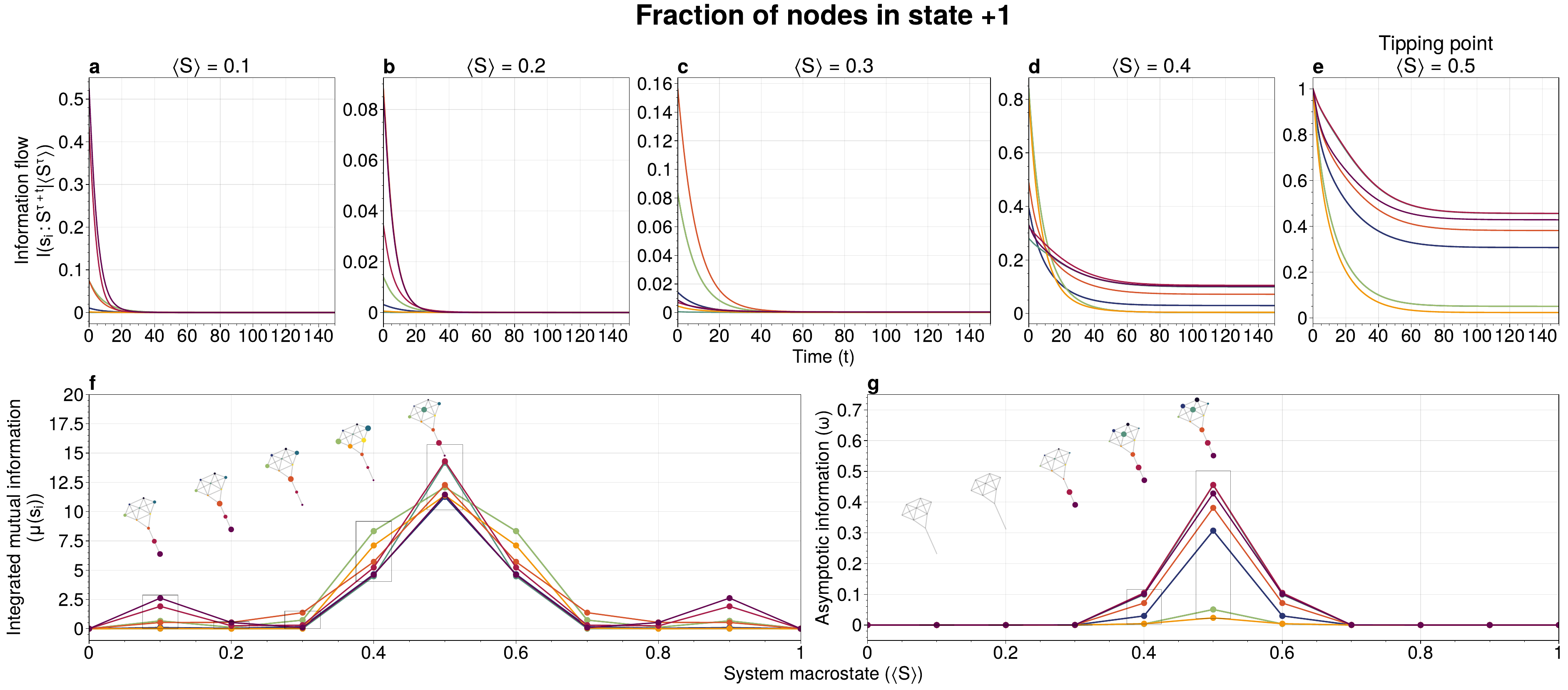}
\caption{\label{fig:kite_res}(a-e) Information flows as distance to tipping point. Far away from the tipping point most information processing occurs in low degree nodes (f,g). As the system moves towards the tipping point, the information flows increase and the information flows move towards higher degrees. (f) Integrated mutual information as function of distance to tipping point. The graphical inset plots show how noise is introduced far away from the tipping point in the tail of the kite graph. As the system approaches the tipping point, the local information dynamics move from the tail to the core of the kite. (g) A rise in asymptotic information indicates the system is close to a tipping point. At the tipping point, the decay maximizes as trajectories stabilize into one of the two attractor states.}
\end{figure*}

Our analysis reveals several key insights into the dynamics of metastable transitions and tipping points in complex networks. We observe a distinct \emph{domino effect} where low-degree nodes initiate system destabilization. As the system approaches a tipping point, information flows shift from low-degree to high-degree nodes. We identify a rise in asymptotic information as a potential early warning signal for an impending tipping point. Finally, we uncover a division of roles among nodes, with some acting as \emph{initiators} that propagate perturbations and others as \emph{stabilizers} that influence the system's transition between attractor states.

In \cref{fig:kite_res}, we visualize the information flows at different stages as the system approaches the tipping point. While we present detailed analysis using the kite graph for simplicity, these findings generalize to other network structures, as demonstrated in \cref{fig:interventions} and further elaborated in the appendix.

\subsection*{Information Flow Dynamics and the Domino Effect}

To decompose the metastable transition, we consider local information flows in a given system partition, \(S_{\gamma} = \{S' \subseteq S | \langle S' \rangle = \gamma\}\) where \(\gamma \in [0,1]\) represents the fraction of nodes having state 1. This yields the conditional integrated mutual information:

\begin{equation}
\label{eq:adj_imi_conditional}
\mu(s_i | \langle S \rangle) = \sum_{t = 0}^\infty (I(s_i^{\tau} : S^{\tau + t} | \langle S^{\tau} \rangle) - \omega_{s_i}) \Delta t.
\end{equation}

Details about the estimation procedure can be found in Appendix \ref{sec:org59af222}.

Two key observations emerge from \cref{fig:kite_res}:

Firstly, the tipping point is reached through a domino effect, with low-degree nodes acting as initiators early in the process. These nodes, being more susceptible to noise (see \cref{fig:maj_flip}), are more likely to pass fluctuations to neighbors -- akin to pushing a ball up a hill.  Far from the tipping point (\cref{fig:kite_res}a), lower-degree nodes show higher integrated mutual information, $\mu(s_i | \langle S \rangle)$, than higher-degree nodes. This noise injection by lower-degree nodes increases the likelihood of a metastable transition.

Secondly, an increase in asymptotic behavior corresponds to the system transitioning between attractor states. As shown in \cref{fig:kite_res}(b, c), asymptotic information remains low far from the tipping point and steadily increases as the system approaches it. Nodes with higher asymptotic information possess greater predictive power regarding which side of the tipping point the system will settle on.

\subsection*{Path Analysis and Tipping Point Trajectories}

To illustrate the information encoded in these flows, we computed trajectories from the attractor state \(S = \{0, \dots, 0\}\), simulated for \(t=5\) steps. \Cref{fig:max_trajectory} shows a trajectory that maximizes:

\begin{equation*}
\label{eq:max_trajectory}
\log p\big(S^{t + 1}|S^{t}, S^0 = \{0, \dots, 0\}, \langle S^5 \rangle = 0.5\big).
\end{equation*}

These trajectories reveal how the information flows measured in \cref{fig:kite_res}c are generated by the sequence of flips originating from the tail of the kite graph. Tail nodes are uniquely positioned to pass on fluctuations to their neighbors, eventually causing a cascade of flips that reach the tipping point. This simple example illustrates how the network structure can influence the system's dynamics and the information flows that precede a metastable transition. Where noise pushes the system towards a tipping point, originating first in low degree nodes for dynamics governed by the Boltzmann-Gibbs distribution.

\subsection*{Network Structure and Node Roles in Metastable Transitions}

The domino effect is not solely determined by node degree. As the system nears the tipping point, network effects become significant. For instance, in the kite graph, node 8 (degree 2) exhibits the highest integrated mutual information when 2 bits are flipped (\cref{fig:kite_res}b). In contrast, node 3 (degree 6) shows low shared information prior to the tipping point but high shared information at the tipping point.

This transition highlights how the network structure as a whole contributes to a system's behavior. Local structural measures, such as degree centrality, may undervalue a node's contribution towards a tipping point and the eventual settlement in a new attractor.

\subsection*{Tipping Point Dynamics and Information Flow}

\begin{figure*}
\centering
\includegraphics[width=.9\linewidth]{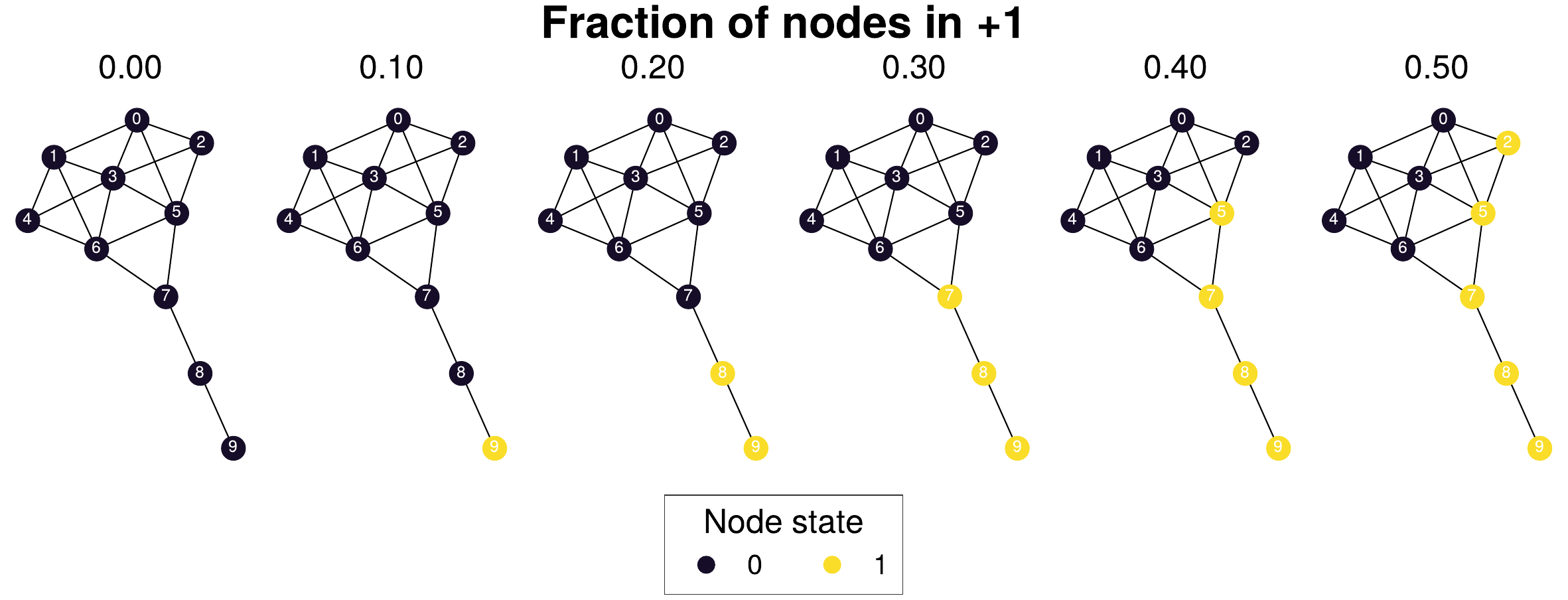}
\caption{\label{fig:max_trajectory}The tipping point is initiated from the bottom up. Each node is colored according to state 0 (black) and state 1 (yellow) Shown is a trajectory towards  the tipping point that maximizes \(\sum_{{t=1}}^{{5}} \log p(S^{{t+1}} | S^t, S^0 =\{0\}, \langle S^5 \rangle ) = 0.5)\). As the system approaches the tipping point, low degree nodes flip first, and recruit ``higher'' degree nodes to further destabilize the system and push it towards a tipping point. In total 30240 trajectories that reach the tipping point in 5 steps, and there are 10 trajectories that have the same maximized values as the trajectory shown in this figure.}
\end{figure*}

\begin{figure*}
\centering
\includegraphics[width=.9\linewidth]{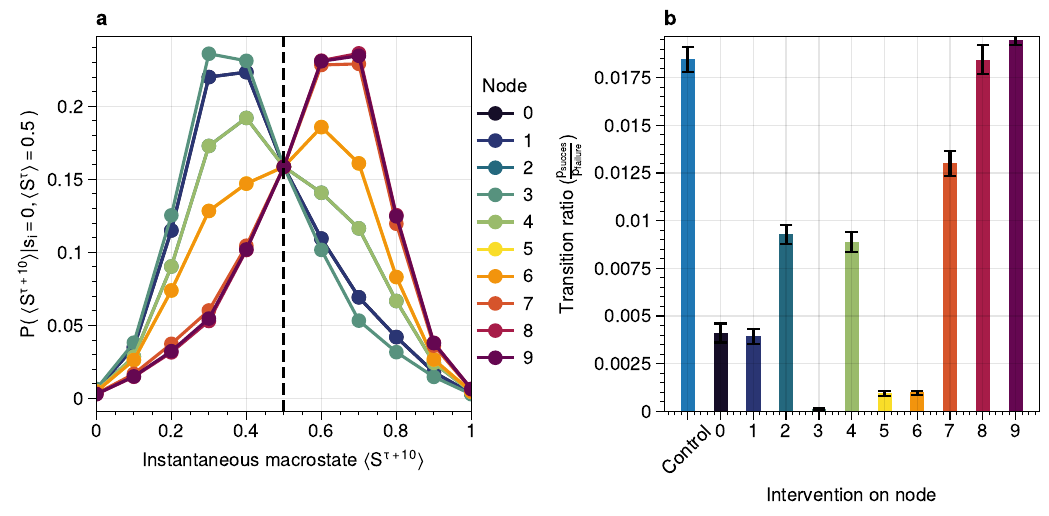}
\caption{\label{fig:butterfly}(a) Shown are the conditional probabilities at time \(t=10\) relative to the tipping point. The shared information between the hub node 3 and the tail node 8 is similar but importantly caused through different sources. The hub (node 3) has high certainty on that the system macrostate will be the same sign as its state. In contrast, node 8 has high certainty that the system macrostate will be opposite to its state at the tipping point. This is caused by the interaction between the network structure and the system dynamics whereby the most likely trajectories to the tipping point from the stable regime is mediated by the noise-induced dynamics from the tail to the core in the kite graph (see main text).(b) Successful metastable transitions are affected by network structure. Successful metastable transitions are those for which the sign of the macrostate is not the same prior and after the tipping point, e.g. the system going from the 0 macrostate side to the +1 macrostate side or vice versa. Shown here are the number of successful metastable transitions for \cref{fig:interventions} under control and pinning interventions on the nodes in the kite graph.}
\end{figure*}

At the tipping point, the system is most likely to either move to a new attractor state or relax back to its original state (\cref{fig:max_trajectory}). Path analysis reveals that the most likely paths to the tipping point result in a configuration where a high-degree cluster of nodes must flip. This trajectory is less likely than reversing the path shown in \cref{fig:max_trajectory}, explaining why most tipping points "fail" and relax back to the original attractor state (\cref{fig:butterfly}b).

The increased information of node 8 around the tipping point can be understood by considering its predictive power about the system's future. As shown in \cref{fig:butterfly}a, both node 3 and node 8 have low uncertainty about the future system state, but the nature of this certainty differs. Node 3 is more certain that the average system state will equal its state at the tipping point, while node 8 is more certain that the future system state will have the opposite sign to its state at the tipping point.

\subsection*{Role Division and Interventions in Tipping Behavior}\label{sec:roles}

We approximate the role of a node \(i\) using the difference between integrated mutual information and asymptotic information:

\begin{equation}
r_i = \max_{\langle S \rangle} \mu^*(s_i | \langle S \rangle) - \max_{\langle S \rangle} \omega^*(s_i) \in [-1, 1],
\end{equation}

where $\mu^*$ and $\omega^*$ are normalized versions of $\mu$ and $\omega$, respectively.

Nodes with role values close to 1 are classified as "initiators," with high predictive information about short-lived system trajectories. Nodes with values close to -1 are "stabilizers," with high long-term predictive information about future system states.

\begin{figure*}
\centering
\includegraphics[width=.9\linewidth]{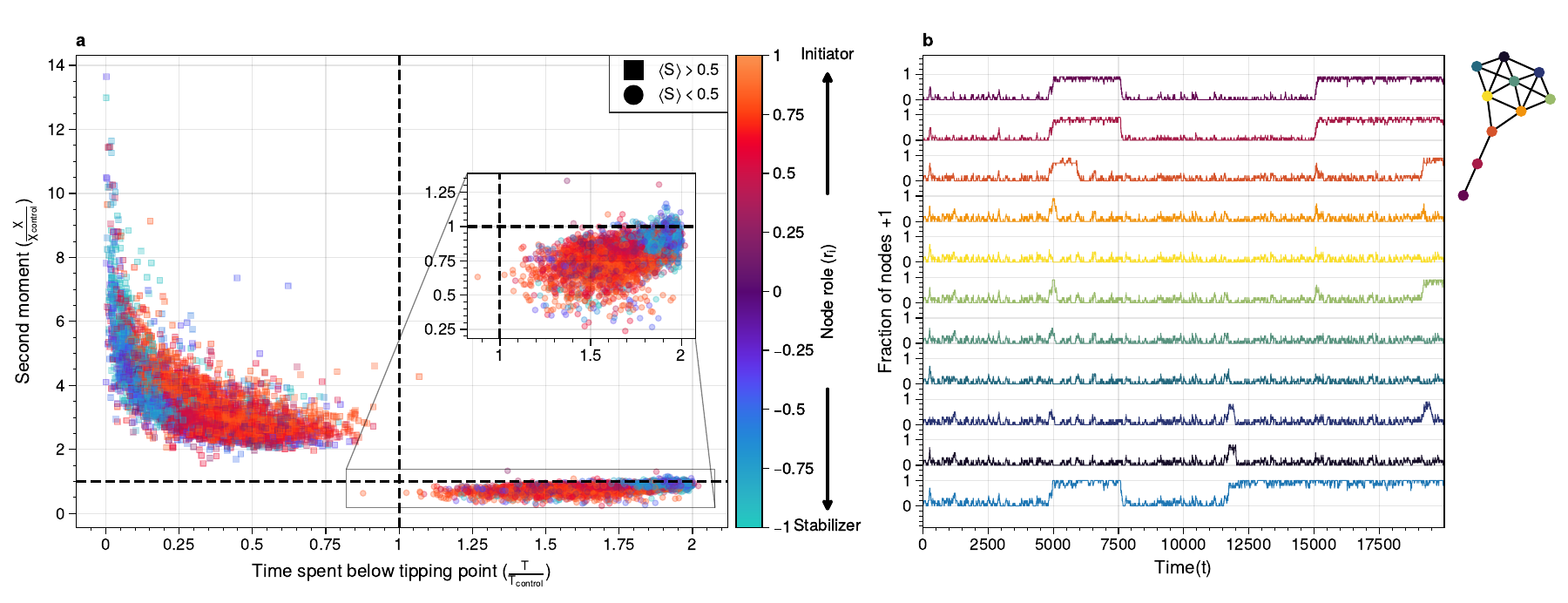}
\caption{\label{fig:interventions}For a system to cross a tipping point, two distinct types of nodes are essential: \textbf{stabilizers}, which contain information about the system's next attractor state and facilitate transitions between states; and \textbf{initiators}, which propagate noise through the system. (a) The effect of causal pinning interventions on node 0 states in Erdos-Renyi graphs ($N = 100$, 10 nodes each, $p = 0.2$, 6 seeds) is shown. Normalized system fluctuations (second moment) and time spent below the tipping point relative to the control are presented per network to indicate the effect of the pinning interventions. Pinning initiators increases tipping points, while pinning stabilizers prevents tipping and increases noise above the tipping point. For more details on role approximation, see \cref{sec:roles}. (b) To exemplify the effect of the causal interventions in (a) typical system trajectories under pinning interventions on a node for the kite graph are shown. Colors reflect intervention on corresponding nodes in the inset kite graph. Initiator-based interventions remove fluctuations below the tipping point ($<0.5$) and increase fluctuations above, whereas stabilizer-based interventions stabilize tipping points while increasing noise.}
\end{figure*}

We validated these roles using simulated interventions (\cref{fig:interventions}). Pinning initiator nodes to the 0 state promotes tipping points, while pinning stabilizer nodes is essential for stabilizing transitions between attractor states.

\section{Discussion}
\label{sec:org389dbab}
Understanding how metastable transitions occur may help in understanding how, for example, a pandemic occurs, or a system undergoes critical failure. In this paper, dynamical networks governed by the Boltzmann-Gibbs distribution were used to study how endogenously generated metastable transitions occur. The external noise parameter (temperature) was fixed such that the statistical complexity of the system behavior was maximized (see appendix \cref{method:appendix}).

The results show that in the network two distinct node types could be identified: initiator and stabilizer nodes. Initiator nodes are essential early in the metastable transition. Due to their high degree of freedom, these nodes are more effected by external noise. They are instigators and propagate noise in the system, destabilizing more stable nodes. In contrast, stabilizer nodes have low degree of freedom and require more energy to change state. These nodes are essential for the metastable behavior as they stabilize the system macrostate. During the metastable transition a domino sequence of node state changes are propagated in an ordered sequence towards the tipping point.

This domino effect was revealed through two information features unveiling an information cascade underpinning the trajectories towards the tipping point.

Integrated mutual information captured how short-lived correlations are passed on from the initiator nodes. In the stable regime (close to the ground state) low degree nodes drive the system dynamics. Low degree nodes destabilize the system, pushing the system closer to the tipping point. In most cases, the initiator nodes will fail in propagating the noise to their neighbors. On rare occasions, however, the cascade is propagated progressively from low degree, to higher and higher degree. A similar domino mechanism was recently found in climate science \cite{Wunderling2020,Wunderling2021}. Wunderling and colleagues provided a simplified model of the climate system, analyzing how various components contribute to the stability of the climate. They found that interactions generally stabilize the system dynamics. If, however, a metastable transition was initialized, noise was propagated through a similar mechanism as found here. That is, an initializer node propagated noise through the system which created a domino effect that percolated through the system.

An increase in asymptotic information forms an indicator of how close the system is to a tipping point. Close to the ground state, the asymptotic information is low, reflecting how transient noise perturbations are not amplified and the system macrostate relaxes back to the ground state. As the system approaches the tipping point, the asymptotic information increases. As the distance to the ground state increases, the system is more likely to transition between metastable states. After the transition, there remains a longer term correlation. Asymptotic information reflects the long(er) timescale dynamics of the system. This ``rest'' information peaks at the tipping point, as the system chooses its next state.

The information viewpoint uniquely offers an alternative view to understand how metastable transitions are generated by dynamical networks. Two information features were introduced that decompose the metastable transition in sources of high information processing (integrated mutual information) and distance of the system to the tipping point (asymptotic information). A domino effect was revealed, whereby low degree nodes initiate the tipping point, making it more likely for higher degree nodes to tip. On the tipping point, long-term correlations stabilize the system inside the new metastable state. Importantly, the information perspective allows for estimating integrated mutual information directly from data without knowing the mechanisms that drive the tipping behavior. The results highlight how short-lived correlations are essential to initiate the information cascade for crossing a tipping point.

\section{Conclusions}
\label{sec:org7971cd6}
Our  information theoretic  approach  offers an  alternative
view   to  understand   \emph{how}  metastable   transitions  are
generated  by dynamical  networks. Two  information features
were introduced that decompose  the metastable transition in
sources  of high  information processing  (integrated mutual
information) and distance of the system to the tipping point
(asymptotic  information).  A  domino effect  was  revealed,
whereby low degree nodes  initiate the tipping point, making
it  more likely  for  higher  degree nodes  to  tip. On  the
tipping point, long-term  correlations stabilizes the system
inside   the   new   metastable  state.   Importantly,   the
information  perspective  allows for  estimating  integrated
mutual information  directly from  data without  knowing the
mechanisms  that drive  the  tipping  behavior. The  results
highlight  how  short-lived  correlations are  essential  to
initiate  the information  cascade  for  crossing a  tipping
point.

\section{Limitations}
\label{sec:org26f073f}
Integrated mutual  information was  computed based  on exact
information  flows. This  means that  for binary  systems it
requires  to  compute a  transfer  matrix  on the  order  of
\(2^{|S|} \times 2^{|S|}\). This  reduced the present analysis to
smaller  graphs. It  would  be possible  to use  Monte-Carlo
methods   to  estimate   the  information   flows.  However,
\(I(s_i^{\tau}  : S^{\tau  + t})\)  remains expensive  to compute.
When using computational models,  it requires to compute the
conditional and  marginal distributions  which are  on order
\(\mathbb{O}(2^{|S|})\)       and       \(\mathbb{O}(2^{t|S|})\)
respectively. In \cref{section:larger_n}, we give a proof of
principle how the results presented here would generalize to larger
systems.

In addition, the decomposition  of the metastable transition
depends  on the  partition of  the state  space. Information
flows are  in essence statistical dependencies  among random
variables. Here,  the effect  of how  the tipping  point was
reached was studied by partition the average system state in
terms of  number of bits flipped.  This partitioning assumes
that the majority  of states prior to the  tipping point are
reached by having fraction \(c  \in [0, 1]\) bits flipped. The
contribution  of  each  system  state  over  time,  however,
reflects a  distribution of  different states;  reaching the
tipping  point from  the  ground  state 0,  can  be done  at
\(t-2\) prior to tipping by either remaining in 0.4 bits, or
transitioning from 0.3 bits flipped to 0.4 and eventually to
0.5 in  2 time steps.  The effect of these  additional paths
showed marginal effects on the integrated mutual information
and asymptotic information.

Information flows  conditioned on a  partition is a  form of
conditional   mutual   information  \cite{James2016}.   Prior
results   showed  that   conditional  information   produces
synergy, i.e. information that is  only present in the joint
of all variables but cannot be found in any of the subset of
each variable.  Unfortunately, there is no  generally agreed
upon    definition    on     how    to    measure    synergy
\cite{Beer2015,Kolchinsky2022}  and different  estimates exist
that may  over or  underestimate the synergetic  effects. By
partitioning one can create synergy as for a given partition
each spin  has some  additional information about  the other
spins. For example, by taking the states such that \(\langle S \rangle =
0.1\),  each spin  ``knows'' that  the average  of the  system
equals 0.1. This creates shared information among the spins.
Analyses  were  performed  to  estimate  synergy  using  the
redundancy  estimation  \(I_{min}\)\cite{Williams2010}.  Using
this  approach, no  synergy was  measured that  affected the
outcome of this study. However, it should be emphasized that
synergetic effects  may influence the  causal interpretation
of the approach presented here.

A  general class  of  systems was  studied  governed by  the
Boltzmann-Gibbs  distribution.  For practical  purposes  the
kinetic Ising model  was only tested, but  we speculate that
the  results should  hold (in  principle) for  other systems
dictated by  the Boltzmann-Gibbs distribution. We  leave the
extension to other system Hamiltonians for future work.
\section{Competing interests}
\label{sec:orgf7ccc8e}
The authors declare no competing interests.
\section{Funding}
\label{sec:org03a6081}
This  research is  supported by  grant Hyperion  2454972 of  the Dutch  National
Police.

\section{Code and Data availability}
The datasets generated and/or analysed during the current study are available in the \url{https://github.com/cvanelteren/metastability} repository,
\url{https://tinyurl.com/4j4ynm4n}.

\section{References}
\label{sec:org26fe258}
\printbibliography[heading=none]

\import{./}{appendix}
\end{document}

%% file: appendix.tex
\appendix
\section{Appendix}
\label{sec:org854db8e}
\counterwithin{figure}{section}
\subsection{Background, scope \& innovation} \label{sec:orgd888f8c}
Noise  induced  transitions   produces  produces  metastable
behavior that is fundamental  for the functioning of complex
dynamical  systems.  For  example, in  neural  systems,  the
presence   of   noise  increases   information   processing.
Similarly, the  relation between glacial ice  ages and earth
eccentricity has  been shown  to have a  strong correlation.
Metastability manifests itself by means of noise that can be
of two  kinds \cite{Forgoston2018}. External  noise originates
from   events   outside   the   internal   system   dynamics
\cite{Calim2021,Czaplicka2013a}.    Examples    include    the
influence of climate effects,  population growth or a random
noise  source  on a  transmission  line.  External noise  is
commonly modeled  by replacing an external  control or order
parameter  by  a  stochastic  process.  Internal  noise,  in
contrast, is inherent to the  system itself and is caused by
random  interactions   of  elements  of  the   system,  e.g.
individuals  in  a  population,  or  molecules  in  chemical
processes.  Both types  of  noise  can generate  transitions
between one metastable state and another. In this paper, the
metastable behavior is studied  of internal noise in complex
dynamical networks governed by the kinetic Ising dynamics.

The ubiquity of multistability  in complex systems calls for
a   general  framework   to   understand  \emph{how}   metastable
transitions occur.  The diversity of complex  systems can be
captured by an interaction networks that dynamically evolves
over  time. These  dynamics can  be seen  as a  distributive
network of  computational units, where each  unit or element
of the  interaction network  changes it  state based  on the
input it  gets from its local  neighborhood. Lizier proposed
that these proposed that  the dynamic interaction of complex
systems  can  be  understood   by  their  local  information
processing \cite{Lizier2008,Lizier2013,Lizier2018}. Instead of
describing  the dynamics  of the  system in  terms of  their
domain  knowledge such  as  voltage  over distance,  disease
spreading rate,  or climate  conditions, one  can understand
the  dynamics in  terms  of the  \emph{information dynamics}.  In
particular, the  field of information dynamics  is concerned
with describing  the system  behavior along its  capacity to
store   information,   transmit  information,   and   modify
information.  By abstracting  away the  domain details  of a
system  and recasting  the dynamics  in terms  of \emph{how}  the
system  computes  its  next   state,  one  can  capture  the
intrinsic computation a system performs. The system behavior
is  encoded in  terms of  probability, and  the relationship
among  these variables  are explored  using the  language of
information theory \cite{Quax2017}.

Information theory offers profound benefits over traditional
methods  used  in  meta-stability analysis  as  the  methods
developed   are    model-free,   can    capture   non-linear
relationships, can be used  for both discrete and continuous
variables,  and   can  be   estimated  directly   from  data
\cite{Cover2005}. Shannon information  measures such as mutual
information and as well as Fisher information can be used to
study how  much information the system  dynamics shares with
the control parameter \cite{Nicolis2016,Lizier2010}.

Past   research   on   information  flows   and   metastable
transitions  focuses on  methods to  detect the  onset of  a
tipping point \cite{Scheffer2009,Prokopenko2011,Scheffer2001}.
It  often centers  around an  observation that  the system's
ability to  absorb noise reduces  prior to the  system going
through a critical point. This critical slowing down, can be
captured  as  a  statistical   signature  where  the  Fisher
information  peaks  \cite{Eason2014}. However,  these  methods
traditionally use some form of control parameter driving the
system  towards   or  away  from  a   critical  point.  Most
real-world systems  lack such an explicit  control parameter
and  require  different  methods. Furthermore,  detecting  a
tipping  point   does  not   necessarily  lead   to  further
understanding  how  the  tipping   point  was  created.  For
example, for a finite size  Ising model, the system produces
bistable behavior. As one increases the noise parameter, the
bistable   behavior  disappears.   The  increase   in  noise
effectively  changes   the  energy  landscape,   but  little
information  is gained  as to  how initially  the metastable
behavior emerged.

In this work,  a novel approach using  information theory is
explored  to  study  metastable  behavior.  The  statistical
coherence between parts of the  system are quantified by the
the  capability of  individual nodes  to predict  the future
behavior  of the  system \cite{Lizier2013}.  Two information
features    are    introduced.    \emph{Integrated    mutual
  information} measure  predictive information of a  node on
the  future  of  the  system.  \emph{Asymptotic  information
  measures} the  long timescale  memory capacity of  a node.
These measures differ from previous information methods such
as  transfer  entropy \cite{Schreiber},  conditional  mutual
information   under   causal   intervention   \cite{Ay2008},
causation   entropy   \cite{Runge2019},   and   time-delayed
variants  \cite{Li2008} in  that these  methods are  used to
infer the transfer  of information between sets  of nodes by
possible correcting for a  third variable. Here, instead, we
aim to understand how the  elements in the system contribute
to the macroscopic properties of the system. It is important
to  emphasize  that  information   flows  are  not  directly
comparable to causal flows \cite{James2016}. A rule of thumb
is  that  causal flows  focus  on  micro-level dynamics  $X$
causes  $Y$,   whereas  information   flows  focus   on  the
predictive aspects,  a holistic view of  emergent structures
\cite{Lizier2013}. In  this sense,  this work is  similar to
predictive  information  \cite{Bialek1999} where  predictive
information  of  some system  $(S)$  is  projected onto  its
consistent elements $(s_i  \in S)$ and computed  as a function
of time $(t)$.

\subsection{Methods and definitions} \label{method:appendix}
\subsubsection{Model}
\label{sec:org5382bb5}
To  study metastable  behavior, we  consider a  system as  a
collection of  random variables \(S =  \{s_1, \dots, s_n\}\)
governed by the Boltzmann-Gibbs distribution

\[p(S)    =     \frac{1}{Z}    \exp(- \beta \mathcal{H}(S) ),\]

where is  the inverse temperature \(\beta  = \frac{1}{T}\) which
control the  noise in the system,  \(\mathcal{H}(S)\) is the
system Hamiltonian which encodes the node-node dynamics. The
choice of the  energy function dictates what  kind of system
behavior we observe. Here, we focus on arguable the simplest
models  that shows  metastable behavior:  the kinetic  Ising
model, and the Susceptible-Infected-Susceptible model.

Temporal  dynamics  are  simulated  using  Glauber  dynamics
sampling.  In each  discrete time  step a  spin is  randomly
chosen  and  a   new  state  \(X'\in  S\)   is  accepted  with
probability

\begin{equation}
\label{eq:glauber}
\begin{split}
 p(  \text{accept} X'  ) =  \frac{1}{1 +
\exp(-\beta   \Delta  E)},
\end{split}
\end{equation}

where  \(\Delta E  =  \mathcal{H}(X') -  \mathcal{H}(X)\) is  the
energy difference  between the  current state \(X\)  and the
proposed state \(X'\).
\subsubsection{Kinetic Ising model}
\label{sec:orgb324012}
The  traditional Ising  model  was  originally developed  to
study ferromagnetism, and is  considered one of the simplest
models that generate complex behavior.  It consists of a set
of binary distributed spins \(S = \{s_1, \dots s_n\}\). Each
spin contains energy given by the Hamiltonian

\begin{equation}
\label{eq:energy}
\begin{split}
\mathcal{H}(S) = -\sum_{i,j} J_{ij} s_{i} s_{j} - h_{i} s_{i}.
\end{split}
\end{equation}

where  \(J_{ij}\) is  the  interaction energy  of the  spins
\(s_i, s_j\).

The  interaction energy  effectively encodes  the underlying
network   structure  of   the   system.  Different   network
structures are used in this study to provide a comprehensive
numerical overview of the relation between network structure
and  information   flows  (see  \cref{method:appendix}).  The
interaction energy  \(J_{ij}\) is set  to 1 if  a connection
exists in the network.

For sufficiently  low noise  (temperature), the  Ising model
shows   metastable  behavior   (\cref{fig:introduction}{c}).
Here,  we aim  to  study  \emph{how} the  system  goes through  a
tipping point by tracking the information flow per node with
the entire system state.
\subsection{Information flow on complex networks}
\label{sec:org3d3e541}
Informally, the information flows measures the statistical coherence
between two random variables \(X\) and \(Y\) over time such that the
present information in \(Y\) cannot be explained by the past of \(Y\)
but rather by the past of \(X\). Estimating information flow is
inherently difficult due to the presence of confounding which potential
traps the interpretation in the ``correlation does not equal causation''.
Under some context, however, information flow can be interpreted as
causal \cite{vanElteren2022}. Let \(S=\{s_1, \dots, s_n\}\) be a random
process, and \(S^t\) represent the state of the random process at some
time \(t\). The information present in \(S\) is given as the Shannon
entropy

\begin{equation}
\label{eq:entropy}
\begin{split}
H(S) = -\sum_{x \in S} p(x) \log p(x)
\end{split}
\end{equation}

where \(\log\) is base 2 unless otherwise stated, and \(p(x)\) is used
as a short-hand for \(p(S  = x)\). Shannon entropy captures the
uncertainty of a random variable; it can be understood as the number of
yes/no questions needed to determine the state of \(S\). This measure of
uncertainty naturally extends to two variables with Shannon mutual
information. Let \(s_i\) be an element of the state of \(S\), then the
Shannon mutual information \(I(S; s_i)\) is given as

\begin{equation}
\label{eq:mi}
\begin{split}
I(S; s_i) &= \sum_{S_i\in S, s' \in s_i} p(S_i,s') \log \frac{p(S_i,s')}{p(S_i)p(s')}\\
          &= H(S) - H(S | s_i)
\end{split}
\end{equation}

Shannon mutual information can be interpreted as the uncertainty
reduction of \(S\) after knowing the state of \(s_i\). Consequently, it
encodes how much statistical coherence \(s_i\) and \(S\) share. Shannon
mutual information can be measured over time to encode how much
\emph{information}  (in bits)  flows  from  state \(s_i^{\tau}\)  to
\(S^{\tau + t}\)

\begin{equation}
\label{eq:flow}
\begin{split}
I(S^{\tau + t}; s_i^{\tau}) = H(S^{\tau + t}) - H(S^{\tau + t} | s_i^{\tau}).
\end{split}
\end{equation}

Prior results showed that the  nodes with the highest causal
importance are those nodes that have the highest information
flow    (i.e.    maximize   \cref{eq:flow})    \cite{vanElteren2022}.
Intuitively,  the   nodes  for   which  the   future  system
``remembers'' information from a node  in the past, is the one
that ``drives''  the system  dynamics. Formally,  these driver
nodes can  be identified by computing  the total information
flow between  \(S^t\) and \(s_i\)  can be captured  with the
integrated mutual information \cite{vanElteren2022}

\begin{equation}
\label{eq:imi}
\begin{split}
\mu(s_i) = \sum_{\tau = 0}^{\infty} I(s_{i}^{t-\tau} ; S^t).
\end{split}
\end{equation}

In some  context, the nodes that  maximizes the \eqref{eq:imi}
are those  nodes that have  the highest causal  influence in
the   system   \cite{vanElteren2022}.   However   in   general
information flows  are difficult  to equate to  causal flows
\cite{Lizier2013,James2016}. Here, the local information flows
are   computed   by   considering  the   integrated   mutual
information conditioned  on part of the  entire state space.
This allows for mapping  the local information flows between
nodes and the system over  time, but does not guarantee that
the measured information flows are directly causal. The main
reason being that having  predictive power about the future,
could  be   completely  caused   by  the   partitioning.  In
\cite{vanElteren2022} the correlation  measured considered all
possible states, and the measures were directly related to a
causal  effect.

In addition,  in \cite{vanElteren2022} the  shared information
between   the  system   with  a   node  shifted   over  time
(\(I(S^{\tau} :  s_i^{\tau + t})\)) was  considered. Applying this
approach under a state partition \(I(S^{\tau} : s_i^{\tau + t} | \langle
S  \rangle)\)causes   a  violation  of  the   data  processing  as
information may flow  from a node at a particular  \(t = t_1\)
and then flow  back to the node  at \(t = t2, t_2  > t_1\). In
order  to simplify  the  interpretation  of the  information
flows and  keep the data processing  inequality, the reverse
\(I(S^{t  + \tau}  : s_i^{\tau}  | \langle  S \rangle)\)  was computed  in the
present study.
\subsection{Noise matching procedure}
\label{sec:org11ee4e3}
The Boltzmann-Gibbs distribution is parameterized by noise factor
\(\beta =  \frac{1}{kT}\) where \(T\) is the temperature and \(k\) is
the Boltzmann constant. For high \(\beta\) values metastable behavior
occurs in the kinetic Ising model. The temperature was chosen such that
the statistical complexity \cite{Lopez-Ruiz1995a} was maximized. The
statistical complexity \(C\) is computed as

\[C = \bar H(S) D(S),\]

\noindent where \(\bar H(S) = \frac{H(s)}{-\log_2(|S|)}\) is the system entropy,
and \(D(S)\) measures the distance to disequilibrium

\[D(S) = \sum_i (p(S_i) - \frac{1}{|S|})^2.\]

\begin{figure}[htbp]
\centering
\includegraphics[width=.9\linewidth]{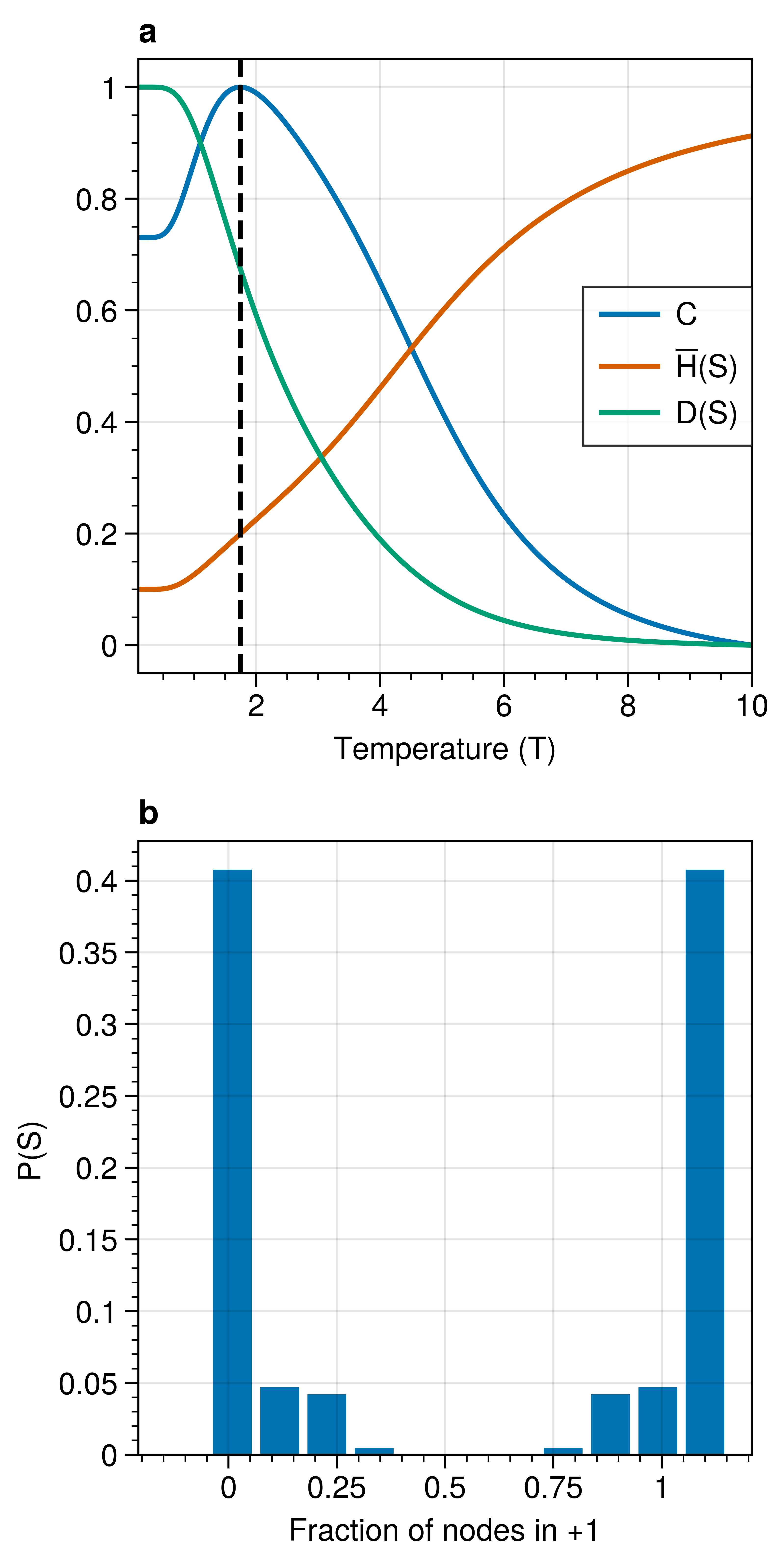}
\caption{\label{fig:stat_compl}(a) Statistical complexity (\(C\)), normalized system entropy (\(H(S)\)) and disequilibrium (\(D(S)\)) as a function of the temperature (\(T = \frac{1}{\beta}\)) for Krackhardt kite graph. The noise parameter was set such that it maximizes the statistical complexity (vertical black line). The values are normalized between [0,1] for aesthetic purposes. (b) State distribution \(p(S)\) for temperature that maximizes the statistical complexity in (a) as a function of nodes in state 1.}
\end{figure}

A typical statistical complexity curve is seen in
\cref{fig:stat_compl}. The noise parameter \(\beta\) is set such that
it maximizes the statistical complexity using numerical optimization
(COBYLA method in scipy's \texttt{optimize.minimize} module)
\cite{Virtanen2020}.

\subsection{Exact information flows \(I(s_i^{\tau} ; S^{\tau + t})\)}
\label{sec:org59af222}
In  order  to  compute  \(I(s_i^{\tau}  :  S^{\tau + t})\),  the
conditional  distribution \(p(S^{\tau  +  t}  | s_i^{\tau})\)  and
\(p(S^{\tau + t})\) needs to  be computed. For Glauber dynamics,
the system  \(S\) transitions into \(S'\)  by considering to
flips  by randomly  choosing  node  \(s_i\). The  transition
matrix \(p(S^t |  s_i) = \textbf{P}\) can  be constructed by
computing each entry \(p_{ij}\) as

\[\label{eq:glauber2}
\begin{split}
p_{ij, i \neq j} &= \frac{1}{|S|} \frac{1}{ 1 + \exp (-\Delta E) }, \textrm{with}\\
p_{ii} &= 1 - \sum_{j, j \neq i} p_{ij},
\end{split}\]

\noindent where \(\Delta E =  \mathcal{H}(S_j) - \mathcal{H}(S_i)\) encodes
the energy difference of moving from \(S_i\) to \(S_j\). The
state to  state transition \(\textbf{P}\) matrix  will be of
size  \(2^{|S|}  \times  2^{|S|} \times  |\mathcal{A}_{s_i}|\),  where
\(|\mathcal{A}_{s_i}|\)  is  the  size of  the  alphabet  of
\(s_i\),  which becomes  computationally intractable  due to
its  exponential growth  with the  system size  \(|S|\). The
exact information  flows can then be  computed by evaluating
\(p(S^t  |  s_i)\)  out  of equilibrium  by  evaluating  all
\(S^t\)  for   all  possible   node  states   \(s_i\)  where
\(p(S^t)\) is computed as

\[p(S^{\tau + t}) = \sum_{s_i} p(S^{\tau + t} | s_i^{\tau} ) p(s_i^{\tau}).\]
\subsubsection{Extrapolation with regressions}
\label{sec:orgf177769}
Exact information  flows were  computed per  graph for  \(t =
500\)  times steps.  Using  ordinary least  squares a  double
exponential was  fit to  estimate the information  flows for
longer \(t\)  and estimate  the integrated  mutual information
and asymptotic information.
\subsection{Noise estimation procedure}
\label{sec:orgc093508}
Tipping point behavior under intervention was quantified by evaluating
the level of noise on both side of the tipping point. Let \(T1\)
represent the ground state where all spins are 0, \(T2\) where all
spins, and the tipping point \(TP\) is where the instantaneous
macrostate \(M(S^t) = 0.5\). Fluctuations of the system macrostate was
evaluated by analyzing the second moment above and below the tipping
point. This was achieved by numerically simulating the system
trajectories under 6 different seeds for \(t = 10e6\) time-steps. The
data was split between two sets (above and below the tipping point) and
the noise \(\eta\) was computed as

\begin{equation*}
\label{eq:noise}
\begin{split}
\eta = \frac{1}{\alpha^2 |S_{w}|}  \sum_w {S_w^t}^2,
\end{split}
\end{equation*}

where \(w \in \{\langle S \rangle < 0.5,\langle S \rangle > 0.5\}\), and

\begin{equation}
\label{eq:noise_estimation}
S_{w}^{t} = \Bigl\{\begin{aligned}
    S^t & \textrm{ if } S^t < 0.5 \\
    1 - S^t & \textrm{ if } S^t > 0.5
    \end{aligned}
\end{equation}

is the instantaneous system trajectory for the system macrostate above
or below the tipping point value. The factor \(\alpha\) corrects for the
reduced range the system macrostate has under interventions. For example
pinning a node \(s_i\) to state 0, reduces the maximum possible
macrostate to \(1 - \frac{1}{n}\) where \(n\) is the size of the system.
The correction factor \(\alpha\) is set such that for an intervention on
0 for a particular node, the range \(S_{\langle S \rangle > 0.5}\)
alpha is set to \(\frac{n}{2} - \frac{1}{n}\).
\subsection{Switch susceptibility as a function of degree}
\label{sec:org009e10c}
First, we investigate the susceptibility of a spin as a function of its
degree. The susceptibility of a spin switching its state is a function
both of the system temperature \(T\) and the system dynamics. The system
dynamics would contribute to the susceptibility through the underlying
network structure either directly or indirectly. The network structure
produces local correlations which affects the switch probability for a
given spin.

As an initial approximation, we consider the susceptibility of a target
spin \(s_i\) to flip from a majority state to a minority state given the
state of its neighbors where the neighbors are not connected among
themselves. Further, the assumption is that for the instantaneous update
of \(s_i\) the configuration of the neighborhood of \(s_i\) can be
considered as the outcome of a binomial trial. Let, \(N\) be a random
variable with state space \(\{0,  1\}^{|N|}\), and let \(n_j \in N\)
represent a neighbor of \(s_i\). We assume that all neighbors of \(s_i\)
are i.i.d. distributed given the instantaneous system magnetization

\[M(S^t) = \frac{1}{|S^t|} \sum_i s_i^t.\]

Let the minority state be 1 and the majority state be 0, the expectation
of \(s_i\) flipping from the majority state to the minority state is
given as:

\begin{dmath}[compact=-1000] \label{eq:majority_flip}
E[ p(s_i = 1 | N ) ]_{p(N)} = \sum_{N_i \in N} p(N_i) p(s_i = 1 | N_i)\\
            = \sum_{N_i \in  N} \prod_j^{|N_i|} p(n_j) p(s_i  = 1 |N_i)\\
            =  \sum_{N_i \in N}  {n\choose k} f^k  (1  - f)^{n-k}  p(s_i  = 1 | f),
\end{dmath}
\vspace{-1em}
where \(f\) is the fraction of nodes in the majority states, \(n\) is
the number of neighbors, \(k\) is the number of nodes in state 0. In
\cref{fig:maj_flip}. This is computed as a function
of the degree of spin \(s_i\). As the degree increases, the
susceptibility for a spin decreases relatively to the same spin with a
lower degree. This implies that the susceptibility of change to random
fluctuations are more likely to occur in nodes with less external
constraints as measured by degree.
\subsection{Additional networks}
\label{additional-networks}
\begin{figure}
\centering
\includegraphics[width=.9\linewidth]{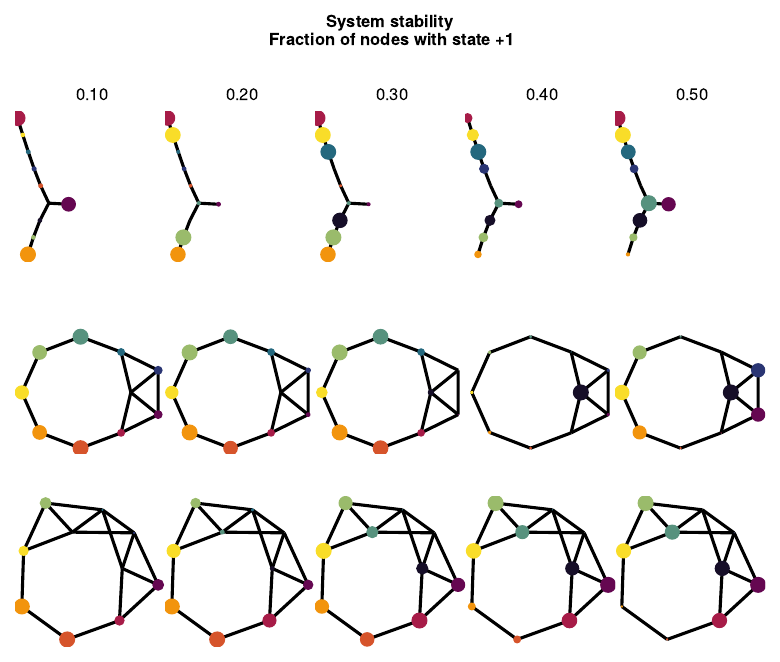}
\caption{\label{fig:other_systems}Adjusted mutual information for a random tree (top), and Leder-Coxeter Fruchte graphs (middle, bottom). Each node is goverened by kinetic Ising spin dyanmics. Far away from the tipping point (fraction nodes 1 = 0.5) most information flows are concentrated on non-hub nodes. As the system approaches the tipping point (fraction = 0.5), the information flows move inwards, generating higher adjusted integrated mutual information for nodes with higher degree.}
\end{figure}

The kite graph was chosen  as it allowed for computing exact
information flows  while retaining a high  variety of degree
distribution given the small  size. Other networks were also
tested.   In  \cref{fig:other_systems}   different  network
structure were used. Each node  is governed by kinetic Ising
spin dynamics.

\subsection{Flip probability per degree}
\label{sec:deg_flip}
\begin{figure}[htbp]
\centering
\includegraphics[width=.9\linewidth]{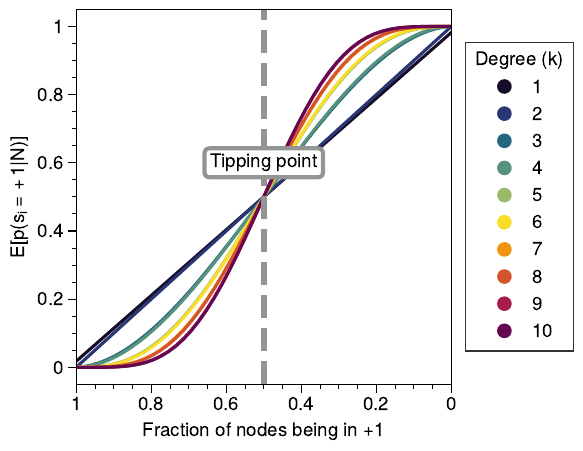}
\caption{\label{fig:maj_flip}Susceptibility of a node with degree \(k\) switching from the minority state 0 to the majority state 1 as a function of the neighborhood entropy for \(\beta = 0.5\). The neighborhood entropy encodes how stable the environment of a spin is. As the system approaches the tipping point, the propensity of a node to flip from to the minority state increases faster for low degree nodes than for high degree nodes. Higher degree nodes require more change in their local environment to flip to the majority state. See for details \cref{sec:org009e10c}.}
\end{figure}

In \cref{fig:maj_flip} the tendency for a node
to flip from the majority  to the minority state is computed
as  function of  fraction of  nodes possessing  the majority
states 1  in the system,  denoted as \(N\). Two  things are
observed.   First,  nodes   with  lower   degree  are   more
susceptible to  noise than nodes  with higher degree.  For a
given system stability, nodes with lower degree tend to have
a higher tendency to flip. This is true for all distances of
the system to the tipping point. In contrast, the higher the
degree of  the node, the  closer the system  has to be  to a
tipping point for the node to  change its state. This can be
explained by  the fact that  lower degree nodes,  have fewer
constraints compared to nodes  with higher degree nodes. For
Ising spin kinetics, the nodes with higher degree tend to be
more ``frozen'' in  their node dynamics than  nodes with lower
degree. Second, in order for a node to flip with probability
with similar  mass, i.e.  (\(E[p(s_i) | N]  = 0.2\))  a node
with higher degree  needs to be closer to  the tipping point
than  nodes  with  lower  degree.  In  fact,  the  order  of
susceptibility   is   correlated   with  the   degree;   the
susceptibility  decreases with  increasing degree  and fixed
fraction of nodes in state 1.

\begin{figure*}[!ht]
\centering
\includegraphics[width=.9\linewidth]{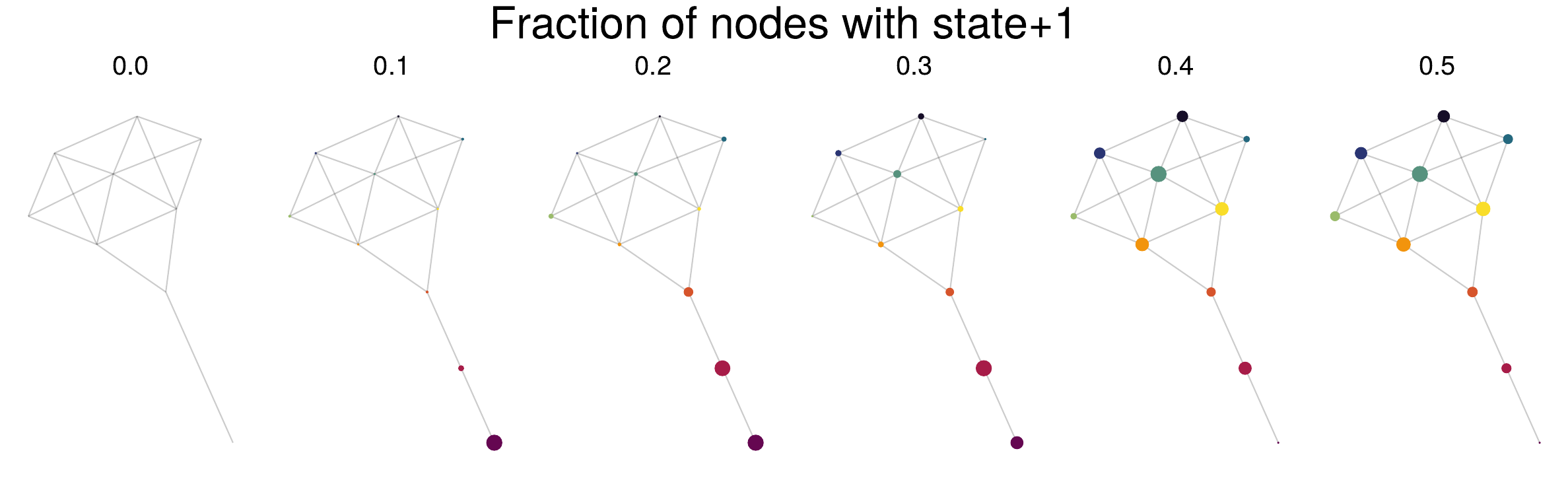}
\caption{\label{fig:expectation_kite}Shortest path analysis of the system ending up in the tipping point from the state where all nodes have state 0. The node size is proportional to the expectation value of a node having state 1  ($E[s_i = 1]_{S^t, M(S^5)}$) as a function of the fraction of nodes having state 1. The expectation values are computed based on 30240 trajectories, an example trajectory can be seen in \cref{fig:max_trajectory}.}
\end{figure*}

\subsection{Synthetic networks}
\label{sec:orgd789b8a}
For  the  synthetic  graphs,  100  non-isomorphic  connected
Erdos-Renyi networks were  generated with a p  = 0.2. Graphs
were generated  randomly and rejected  if the graph  did not
contain a  giant component,  or was isomorphic  with already
generated graphs. For each of the graphs, information curves
were computed as function of the macrostate \(\langle S \rangle\).

\begin{figure}[htbp]
\centering
\includegraphics[width=.9\linewidth]{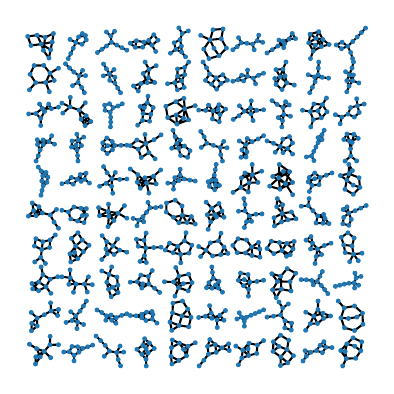}
\caption{\label{fig:ER_family}Erdos-Renyi graphs generated from seed = 0 to produce non-isomorphic connected graphs.}
\end{figure}
\subsubsection{Noise and time spent}
\label{sec:org1eb5d04}
Various network  structures are  generated in  the synthetic
networks. The  variety of  network structure  has non-linear
effects on the information flows. The effect of intervention
in \cref{fig:interventions} is made relative  to the control values for
the  graph and  seed. The  second moment  (appendix: \cref{sec:orgc093508}) and the time spent below the tipping
point   are   normalized   with   respect   to   the   graph
(\cref{fig:ER_family}) and  the seed.  In total 6  seeds are
used (0, 12, 123, 1234, 123456, 1234567).

\subsection{Case Study of a Larger System} \label{section:larger_n}
\begin{figure*}[!htbp]
\centering
\includegraphics[width=0.9 \textwidth]{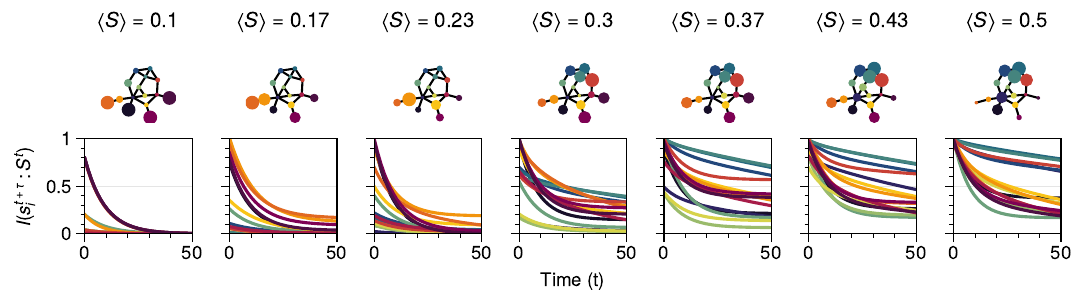}
\caption{Example of tipping behavior in a system consisting of $N = 15$ nodes. The information decay curves are bundled per degree. The transition from left to right increases the number of bits flipped until the tipping point. A wave can be seen where the integrated information from lower-degree nodes to higher ones as the number of bits flipped increases. The size of the nodes are propportional to the integrated mutual information.}
\label{fig:florentine_family}
\end{figure*}

In this section, we extend our analysis to a 15-node network to demonstrate the applicability of our findings to larger systems (see \cref{fig:florentine_family}). This case study serves to validate our theoretical insights derived from smaller networks and to illustrate how the fundamental mechanisms of metastable transitions are preserved as network size increases. Despite the increased computational complexity, our results indicate that the structural features driving these transitions in smaller networks are also evident in larger ones.

\FloatBarrier

As highlighted in the \cref{sec:org26f073f}, the state space of a network grows exponentially ($2^n$) with the number of nodes, making simulations of larger systems computationally demanding. Nevertheless, our analysis of the 15-node network supports our assertion that the foundational processes identified in our primary study can be extrapolated to more complex networks. Detailed results and discussion of this 15-node network analysis are provided
to substantiate our approach and highlight the consistency of our findings across different network sizes.